\documentclass{icrc29}
\usepackage{graphicx,amssymb,amsmath,times}
\begin{document}

\title[Determination of the Aperture of the Pierre Auger Observatory surface detector]{Aperture calculation of the Pierre Auger Observatory surface detector}

\author[Allard \emph{et al.}, for the Pierre Auger Collaboration]{D. Allard, I. Allekotte, E. Armengaud, J. Aublin, X. Bertou, A. Chou, P. L. Ghia,
\newauthor
M. G\'omez Berisso, J.C. Hamilton, I. Lhenry-Yvon, C. Medina, G. Navarra,
\newauthor
E. Parizot, A. Tripathi, for the Pierre Auger Collaboration\\
Observatorio Pierre Auger, Av San Martin Norte 304,(5613) Malargue, Argentina
}
\presenter{Presenter: E. Parizot (parizot@ipno.in2p3.fr), \  fra-parizotE-abs1-he14-poster}
\maketitle

\begin{abstract}
We determine the instantaneous aperture and integrated exposure of the surface detector of the Pierre Auger Observatory, taking into account the trigger efficiency as a function of the energy, arrival direction (with zenith angle lower than 60 degrees) and nature of the primary cosmic-ray. We make use of the so-called Lateral Trigger Probability function (or LTP) associated with an extensive air shower, which summarizes all the relevant information about the physics of the shower, the water tank Cherenkov detector, and the triggers.

\end{abstract}

\section{Introduction}
\label{Intro}

The surface detector (SD) of the Pierre Auger Observatory (PAO) has been continuously growing for the last two years, and the last of its 1600 water tanks is expected to be ``online'' in early 2006. During this intense deployment phase, the SD instantaneous aperture keeps increasing at an irregular pace, depending on the water tank deployment rate and the evolving geometry of the array. While the \emph{local trigger} rate (i.e. at the level of individual stations) is simply proportional to the number of stations in the SD array, the number of exploitable events increases at a rate depending on the \emph{central trigger} conditions and on specific cuts applied to guarantee a high-quality data set. The chain of triggers is described in \cite{Trigger}. Most relevant to the SD acceptance calculations are the so-called high-level triggers, namely the \emph{physics trigger} T4 selecting real cosmic-ray (CR) events in a data set partially contaminated by chance coincidences between neighbour tanks, and the \emph{quality trigger} T5 selecting events offering nominal reconstruction accuracy.

The standard SD data set is composed of events from CR-induced showers developing along a zenith angle $\theta \le 60^{\circ}$. These events are selected very efficiently by the T4 condition requiring that at least 3 tanks pass the time-over-threshold (ToT) trigger (see \cite{Trigger}). The T5 trigger applied to the data set used in this Conference consists of two conditions: 1) the SD station recording the highest signal from the shower must be surrounded by at least 5 active tanks among the 6 closest neighbours, and 2) the reconstructed shower core must be inside an equilateral triangle of active stations at the time of the event. This guarantees that no crucial information is missed for the shower reconstruction: it essentially rejects events located close to the array boundary or close to a temporarily missing tank inside the array.

\section{Detection efficiency in an elementary hexagon}

\subsection{Lateral Trigger Probability function (LTP)}

\begin{figure*}[t]
\hfill\includegraphics[height=6cm]{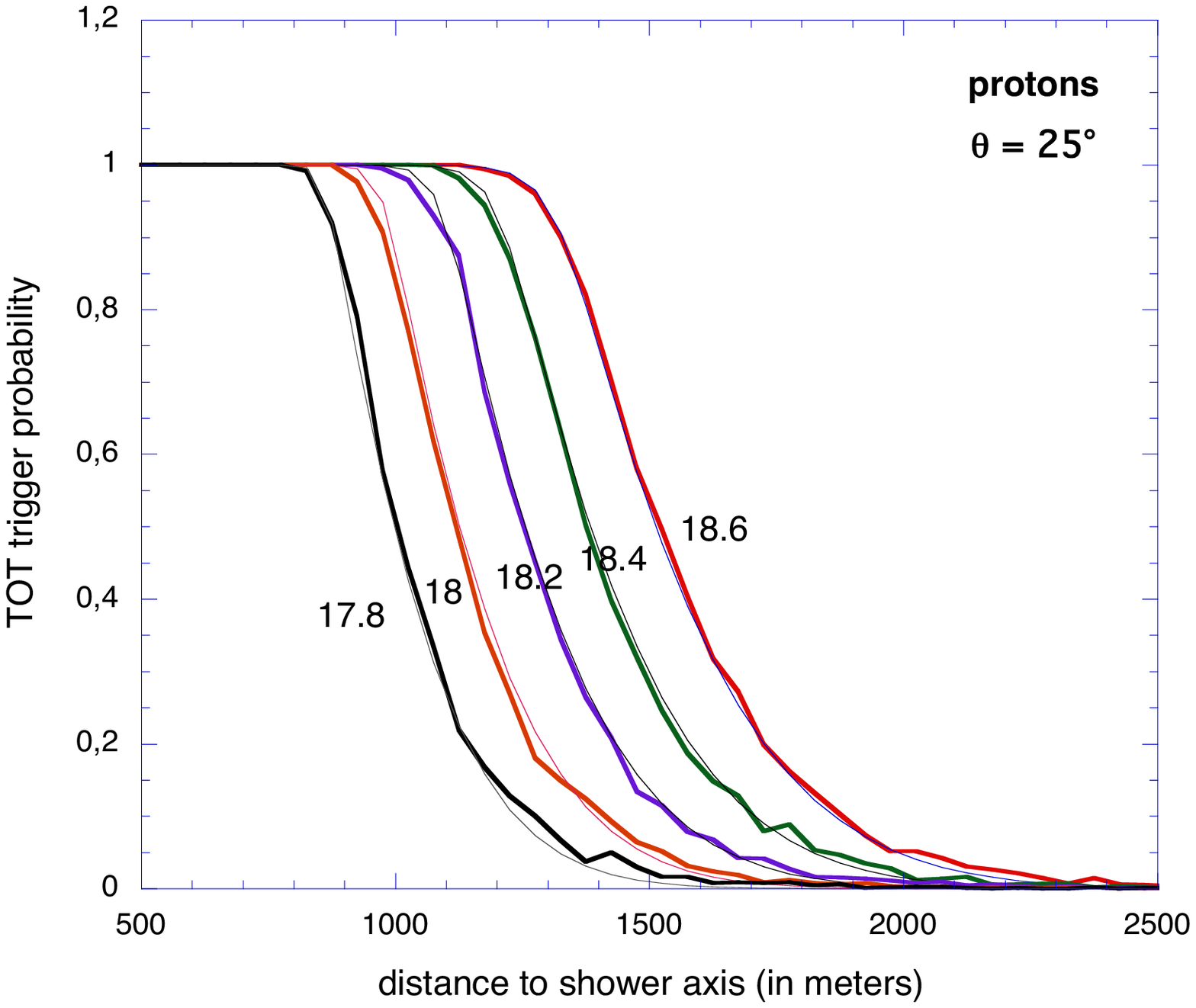}
\hfill\includegraphics[height=6cm]{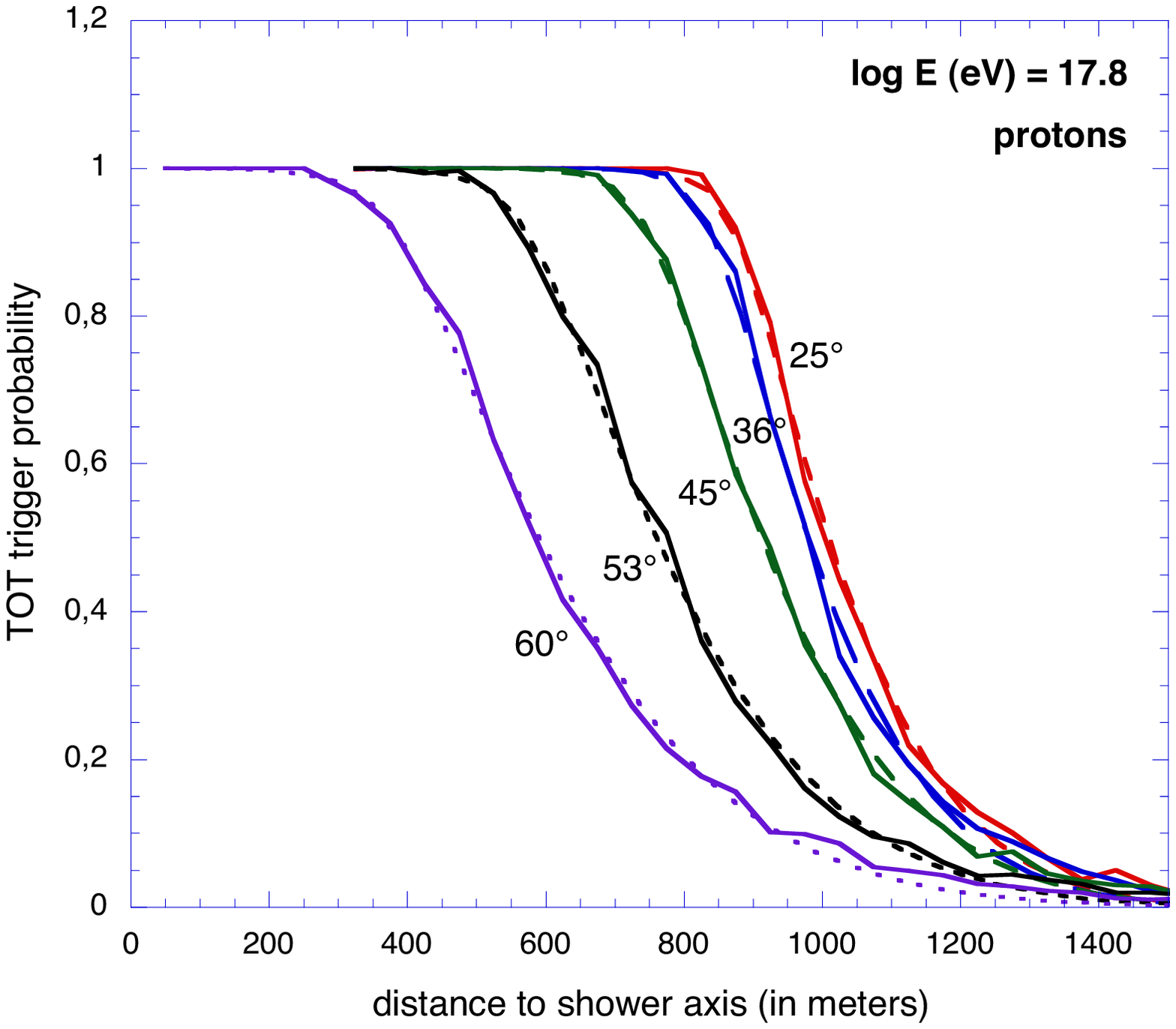}\hfill~
\caption{Examples of LTP functions for proton-induced showers at a zenith angle $\theta = 25^{\circ}$ for different energies labeled by $\log_{10}(E)$, where $E$ is in~eV (left) and at energy $E = 10^{17.8}$~eV for different zenith angles (right).}
\label{fig:LTPs}
\end{figure*}

At high energy, CR-induced showers have a big enough print on the ground to trigger many SD stations independently of where they exactly fall in the array. At lower energy, however, the granularity of the array becomes important and the ability of a given shower to trigger (at least) three tanks depends on the position of the shower core (i.e. the intersection of the shower axis with the ground) with respect to the closest tanks. The particle density in the shower is a steeply decreasing function of the distance to the shower axis, $r_{\mathrm{sf}}$ (radial coordinate away from the axis), and so is the probability for a given shower to trigger a tank.

As recalled above, for the nominal PAO SD data set we use a ``3ToT condition'' and we are thus interested in the probability that a given CR-induced shower triggers at least 3 stations with the ToT condition. It is convenient to introduce the so-called Lateral Trigger Probability function (LTP), giving the \emph{local} (individual tank) ToT trigger probability as function of $r_{\mathrm{sf}}$, for any shower induced by CRs of type $i$ (protons, nuclei, photons...), energy $E$ and arrival direction with zenith angle $\theta$ and azimuth $\phi$. At the energies of interest, the influence of the geomagnetic field and thus the dependence on $\phi$ are negligible. From the practical point of view, the LTP functions can be obtained from a set of Monte-Carlo simulations of the primary CR interaction, shower development and tank response, as the fraction of tanks that pass the trigger in different bins of distance, $r_{\mathrm{sf}}$, from a large number of showers with given parameters: $P_{i,E,\theta}(r_{\mathrm{sf}}) = (\mathrm{No.\,of\,stations\,passing\,the\,trigger\,at}\, r_{\mathrm{sf}})/(\mathrm{total\,No.\,of\,stations\,at\,distance}\,r_{\mathrm{sf}})$.

Examples of LTP functions are shown in Fig.~\ref{fig:LTPs} for proton showers simulated with the Corsika Monte-Carlo program\cite{Corsika} using the QGSJet hadronic model\cite{GQSJet}, for different primary energies and zenith angles. Obviously, the trigger probability is 1 close to the shower core and 0 far away from it. The transition radius, $r_{1/2}$, such that $P_{i,E,\theta}(r_{1/2}) = 1/2$ is an increasing function of energy (as the shower gets bigger) and a decreasing function of $\theta$ (as the shower gets older). The transition region also grows from $\sim 200$~m at 25$^{\circ}$ to $\sim 600$~m at 60$^{\circ}$.

It is worth emphasizing that the LTP functions gather all the information relevant to aperture and exposure calculations, summarizing the properties of the shower development, tank response and trigger definition. To each choice of simulation codes and local trigger definition corresponds a set of LTPs from which the PAO detection efficiency can be computed straightforwardly. When we have larger statistics, the LTPs will be derived directly from the data using hybrid events to measure the SD trigger probability, so that it will be possible to correct for the partial efficiency at low energy in a model-independent way. The systematic comparison of measured and predicted LTPs will also provide valuable information about the physical ingredients of the models.

\begin{figure}[t]
\hfill\includegraphics[height=6cm]{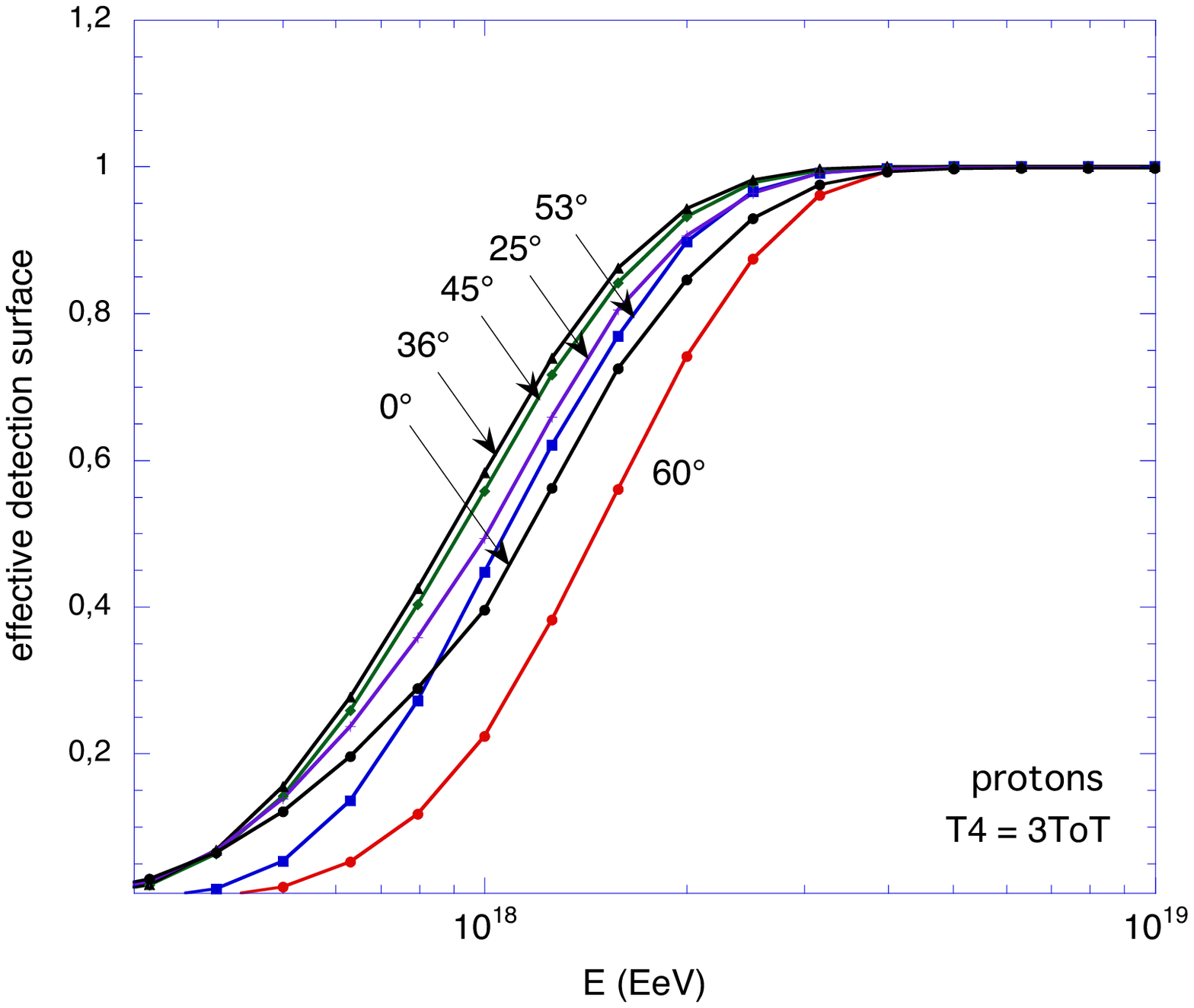}
\hfill\includegraphics[height=6cm]{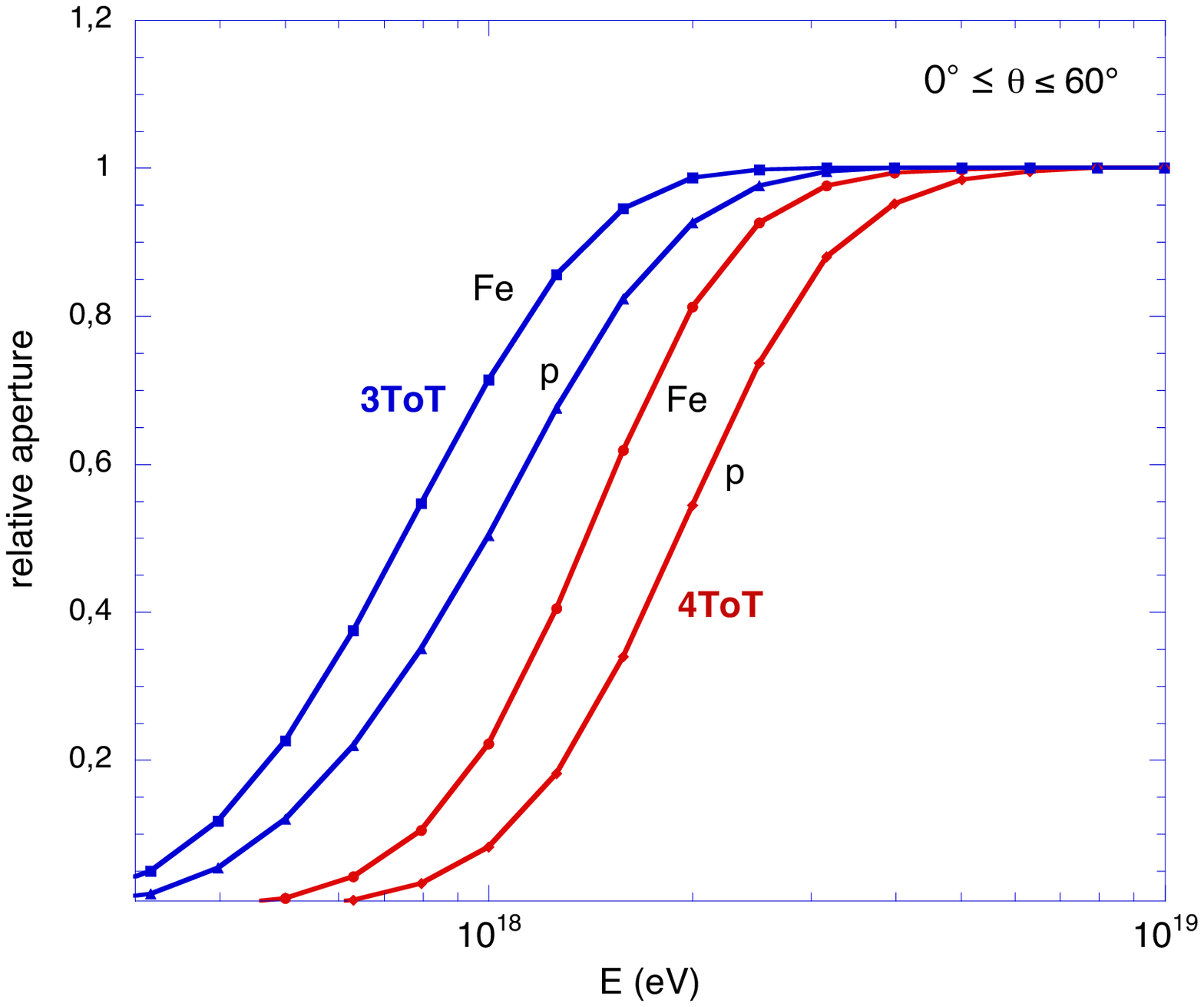}
\caption{Left: Effective detection surface as a function of energy for proton showers at different zenith angles. Right: Aperture saturation curves for proton or Fe-induced showers, with either 3ToT or 4ToT selection criterion (see text).}
\label{fig:effectiveArea}
\end{figure}

\subsection{Effective detection surface and elementary aperture}

For any shower core position and arrival direction, one can compute the distance of each tank in the shower frame and deduce its trigger probability from the above LTP functions. The probability to pass the global T4 trigger then follows from a simple combination of these local LTP values. Let us consider an elementary hexagonal cell of the SD array, with a central station and 6 closest neighbours, labeled by $i\in \{1,...7\}$. For a given shower with core position $(x_{\mathrm{c}},y_{\mathrm{c}})$, let $P_{i}$ and $Q_{i}\equiv 1 - P_{i}$ be the probability that station $i$ triggers and does not trigger, respectively. The probability that none of the 7 tanks gets triggered is simply $\mathcal{P}(0) = \prod_{i=1}^{i=7}Q_{i}$. Likewise, the probability that one and only one tank gets triggered is $\mathcal{P}(1) = \sum_{i}\mathcal{P}(0)\times(P_{i}/Q_{i})$. Finally, the probability that the shower passes the high-level T4 trigger and be part of the PAO data set, i.e. the probability that at least 3 tanks pass the ToT trigger is given by:
\vspace{-10pt}
\begin{equation}
\mathcal{P}_{\mathrm{T4}}(x_{\mathrm{c}},y_{\mathrm{c}}) = 1 - \mathcal{P}(0) \left[1 + \sum_{i}\frac{P_{i}}{Q_{i}} + \sum_{j>i}\frac{P_{i}}{Q_{i}}\frac{P_{j}}{Q_{j}}\right].
\label{eq:3ToTProba}
\end{equation}

\begin{figure}[!h]
\begin{minipage}[t]{0.48\textwidth}
\mbox{}\\
Any tank with six active neighbours (Fig.~\ref{fig:hexagones}a) contributes exactly the elementary hexagon aperture, while any tank with five neighbours contributes 4/6th of that, as illustrated in Fig.~\ref{fig:hexagones}b. Tanks with less than 5 neighbours do not contribute at all. The aperture of a hexagonal cell is thus the building block of all aperture calculations, to be computed once and for all or measured thanks to hybrid data. At full efficiency, i.e. above $\sim 3$~EeV, the detection area per internal tank (with 6 neighbours) is $D^{2}\times\sqrt{3}/2 \simeq 1.95\,\mathrm{km}^{2}$, where $D = 1.5$~km is the array spacing, and the corresponding aperture for showers with $\theta < 60^{\circ}$ is then $a_{\mathrm{cell}} \simeq 4.59\,\mathrm{km}^{2}\,\mathrm{sr}$.
\end{minipage}
\hfill
\begin{minipage}[t]{0.48\textwidth}
\mbox{}\\
\centerline{\includegraphics[width=\linewidth]{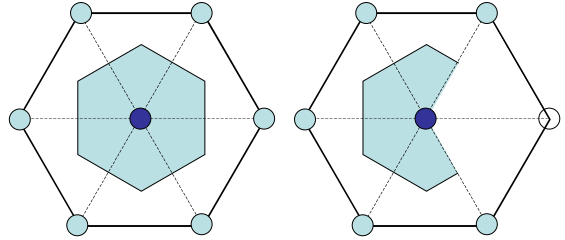}}
\caption{\label{fig:hexagones}Schematic view of the area (shaded region) where the shower core must be located inside an elementary hexagonal cell in order to pass the T5 trigger (left: full hexagon; right: hexagon with a missing vertex).}
\end{minipage}
\end{figure}

\vspace{-6pt}
This allows us to calculate the trigger probability for showers of any energy, mass, and angle as a function of the core position. The global shower detection efficiency is then obtained by averaging over all the allowed shower core positions, i.e. satisfying the T5 condition. They are shown as shaded area in Fig.~\ref{fig:hexagones}. On the left, where all 7 tanks are present, the allowed core positions are such that the closest tank (which is bound to record the largest signal) is the central one, since all the others do not satisfy the requirement of having 5 neighbours. On the right, the rightmost tank is missing and shower core positions in the two concerned triangles are excluded.

In practice, we compute the \emph{effective detection surface} of the elementary hexagonal cell as the integral of its area (in the allowed region) weighted by the trigger probability:
\begin{equation}
S_{\mathrm{eff}} = \int_{\mathrm{cell}} \mathcal{P}_{\mathrm{T4}}(x_{\mathrm{c}},y_{\mathrm{c}}) \times H_{\mathrm{T5}}(x_{\mathrm{c}},y_{\mathrm{c}}) \times \mathrm{d}S,
\label{eq:SEff}
\end{equation}
where $H_{\mathrm{T5}}(x,y)$ implements the T5 condition and is equal to either 1 or 0, depending on whether the core position is allowed or not. Results are shown in Fig.~\ref{fig:effectiveArea}a for the same proton showers as above, as a function of energy: $S_{\mathrm{eff}}$ saturates at $\sim 3$~EeV for moderately inclined showers, and $\sim 4$~EeV for showers at $\theta = 0^{\circ}$ or $60^{\circ}$. Finally, the aperture is obtained by integrating over solid angle: $a(E) = \int_{\theta\le60^{\circ}}S_{\mathrm{eff}}(E,\theta,\phi)\cos\theta\times \sin\theta\mathrm{d}\phi\mathrm{d}\theta$. Results are shown in Fig.~\ref{fig:effectiveArea}b for two different choices of the T4 condition: at least 3 or at least 4 stations with a ToT trigger. The latter thus requires one more tank and is then less efficient at low energy. Figure~\ref{fig:effectiveArea}b also shows that the SD array is more sensitive to Fe nuclei than to protons at low energy, which can be used to constrain the primary composition of CRs in the astrophysically important region of the ankle. The SD detection efficiency is 1 above 3~EeV and 7~EeV for the 3ToT and 4ToT triggers, respectively.

\section{Instantaneous aperture of the SD array and integrated exposure}

The T5 trigger allows us to exploit the regularity of the hexagonal array in a very simple way. The instantaneous aperture of any array configuration is obtained as a multiple of the above-mentioned elementary aperture: we simply need to consider each tank, one after the other, and determine its contribution to the global aperture.

To calculate the integrated exposure of the SD over a given period of time, one finally has to count the number of \emph{cell-seconds} corresponding to that period. The array configuration changes occasionally when new tanks are deployed or when stations experience failures (e.g. electronics, power supply, communications...) and thus get temporarily ``out'' of the array (or back in). These changes in the array configuration are monitored through the local trigger rate with a time resolution of one second (much better than needed for exposure calculations!). For each new configuration, the number of elementary cells, $N_{\mathrm{cell}}$, is computed and the exposure is incremented by $N_{\mathrm{cell}}\times a_{\mathrm{cell}}\times \Delta t$, where $\Delta t$ is the duration of the configuration. For the period from Jan. 1st, 2004 to June 5th, 2005, we obtained $1.21\times 10^{10}$ cell-seconds, which corresponds to 1750~km$^{2}$\,sr\,yr. At low energy, this number must be scaled according to the relative aperture given in Fig.~\ref{fig:effectiveArea}b.

The accuracy of this exposure calculation is expected to be excellent at energies above saturation, since it is based on purely geometrical considerations. However, the monitoring of the instantaneous array configurations may not be complete, and we have considered the possibility of \emph{hidden dead times} that would not appear at the tank level or in the communication chain from the stations to the CDAS. This will be monitored appropriately in the near future. By comparing the daily averaged number of events in our data set with the mean number expected from the exposure calculations (with Poissonian fluctuations), we estimated a possible error of $\sim 5\%$. That would thus lead to an error of $\sim 5\%$ on the differential CR flux, equivalent to an energy shift of less than 2\%, which is negligible compared to the PAO energy accuracy. The uncertainty on the integrated exposure can thus be considered as negligible.

\end{document}